# Defining the Structure of Environmental Competence of Future Mining Engineers: ICT Approach

Vladimir S. Morkun[1][0000-0003-1506-9759], Serhiy O. Semerikov[2][0000-0003-0789-0272],
Nataliya V. Morkun[1][0000-0002-1261-1170], Svitlana M. Hryshchenko[1][0000-0003-4957-0904]
and Arnold E. Kiv[3]

[1] State Institution of Higher Education "Kryvyi Rih National University",
11, Vitali Matusevich St., Kryvyi Rih, 50027, Ukraine
`{morkunv, nmorkun}@gmail.com, s-grischenko@ukr.net`
[2] Kryvyi Rih State Pedagogical University, 54, Gagarina Ave., Kryvyi Rih, 50086 Ukraine
`semerikov@gmail.com`
[3] Ben-Gurion University of the Negev, Beer Sheba, Israel
`kiv@bgu.ac.il`

**Abstract.** The *object* is to the reasonable selection of the ICT tools for formation of ecological competence. Pressing *task* is constructive and research approach to preparation of future engineers to performance of professional duties in order to make them capable to develop engineering projects independently and exercise control competently. *Subject of research*: the theoretical justification of competence system of future mining engineers. *Methods*: source analysis on the problem of ecological competence formation. *Results*: defining the structure of environmental competence of future mining engineers. *Conclusion*: the relevance of the material covered in the article, due to the need to ensure the effectiveness of the educational process in the preparation of the future mining engineers.

## 1 Introduction

Nowadays information and communication technologies (ICT) applying is one of the principle education tasks, which provides the education process improvement, its accessibility and effectiveness and preparing the younger generation for living in the information society. Education must serve for the society needs. Processes reflecting current trends in society provide information technologies development and implementing. At present the ICT usage in education can be a catalyst in solving important social problems connected with increasing the educational resources and services availability and quality, real and equal opportunities in getting education for citizens despite their residence, social status and income [6]. Currently, high technologies cover almost all areas of our life. Professionals having common practical and theoretical skills of work with different information types are highly wanted.

One of the most important education tasks is to develop students' active cognitive attitude to knowledge. Cognitive activity in universities is a necessary stage in preparing for further professional life. Teacher's task is to seek and find the best methods and tools of improving the educational process and leading to the cognitive interest development.

## 2 Materials and methods

The study is carried out in "Kryvyi Rih National University" according to the plan of joint research laboratory using cloud technologies in education process of "Kryvyi Rih National University" and the Institute of Information Technologies within the research project "Adaptive system of individual teaching for mining engineers based on the integrated structure of artificial intelligence "A digital tutor" [8]. The author analyzes sources devoted to investigating the problems of ecological competence formation and usage of geoinformation technology in the teaching future mining engineers [7, 10]. This research also improves the system of competence among future mining engineers, gives its theoretical explanation and represents geoinformation technology means used in education process [11, 12].

## 3 Results and discussion

Prerequisite for development of methodic of usage of geoinformation technologies as a tool for formation of environmental competence of future mining engineers is the solution of the following specific tasks of research:

1. determination of major factors of modernization of professional education of future mining engineer;
2. theoretical justification of competence system of future mining engineers;
3. defining the structure of environmental competence of future mining engineers;
4. earmarking of geoinformation technologies, application of which promotes safe activity of mining plants.

In result of solution of the first task, the major factors for modernization of professional education of future mining engineers is the public contract for preparation of competent specialists were determined. They are specified in the state industry standards of higher education and in the society sustainable development.

The main regulatory document determining the judicial and basic arrangements for the activity of mining engineers concerning mining works performance, securing emergency protection of mining plants, establishments and organizations is the Law on mining of Ukraine (1999).

Abdallah M. Hasna [5] considers the contribution of the principals stated in the article 7 of the Mining law of Ukraine [6] into social, economical, ecological and technical development as the function of sustainable development – model of resource use, focused on the satisfaction of human's wants when preserving of the environment in such a way, that all these wants could be satisfied not only by current generations but also by future ones. In such a way state policy in mining industry focused on the sustainable development of mining industry, science and education.

In result of solution of the second task, it was stated that development of competency building approach to professional education happens, on the one side, under the influence of public contract on the preparation of competent specialists, on the other side – it influences on the formation of such contract in the direction of changing of state industry standards of higher education.

Application of competency building approach to modernization of state industry standards of higher education leads to the necessity of theoretical justification and development of competence system of future mining engineers, the component of which are environmental competences.

In result of solution of the third task it was stated that formation of environmental competence of future mining engineer happens within professional education.

The main environmental requirements in the field of mining works, prevention of ill effect of mining works and securing of ecological safety during mining works is not only a subject matter of certain articles of the Mining law of Ukraine, but also obligate constituent of preparation of environmentally competent mining engineer [13, 14].

In result of solution of the fourth task it was concluded that development of ICT [15] promotes the changes of production technological mode (including mining production), which provides stable technological development, namely geoinformation technologies. Education modernization in today's society can be hardly imagined without education innovation developing especially such as GIS technologies. GIS is an integrated set of hardware, software and media means, providing input, storing, processing, manipulating, image analyzing and space-coordinated data representation [3]. GIS using in education allows to perform independent analyzing, interrelations search, analogy detecting and developing abilities to explain the differences. The last statement is true to engineering professionals. This implementation in the engineering profession also has allowed to reduce waste of time during the session, to create specific pedagogical conditions for developing future professionals skills, increase cognitive interest, to set subjective position in learning activities, to build cognitive autonomy, students' information and communication competence, motivational readiness for cognitive activity. The results of the analysis show that the main key words that different researchers refer to environmental competence, are as follows:

— in relation to the subject activity: "person", "personality", "personal", "education", "characteristics", "ability", "willingness", "quality", "behavior", "society";
— in relation to the object of activity: "value", "moral nature", "environment", "nature", "natural", "preservation";
— in relation to the content and nature of the activity: "environmental", "professional", "practical", "experience", "ability", "skill", "use", "knowledge", "cognitive", "system", "provide", "significant".

Due to the fact that, by definition of DeSeCo specialists [16], environmental sustainability (ecological sustainability) is the basis of the key competences of the individual associated with success in society, consideration of environmental competency is advantageously carried out at three levels:

- on the general level of ecological culture and environmental awareness (Zenobia Barlow and Michael K. Stone [1]);
- on a social-professional level of environmental literacy (Carmel Bofinger [2], B. E. Harvey [4]);
- on the special professional level of environmental competence (Svitlana M. Hryshchenko, Vladimir S. Morkun and Serhii O. Semerikov [9]).

The carried out analysis gives the possibility to determine environmental competence of future mining engineer as personal formation, which includes the acquired during preparation profession-oriented environmental awareness (cognitive component), adopted ways for securing environmentally safe mining works (praxeological component) in the interest of sustainable development (axiological component) and the qualities of socially responsible ecological behavior (socially-behavioral component) are formed [9].

By definition, formation of environmental competency of future mining engineer happens during professional education of bachelors in mining, that is why for determination of environmental competencies we will refer to the components of developed system of socially-personal, instrumental, general scientific, general professional and specially professional competences of future mining engineer.

In their turn the requirements for stable social, economical and environmental society development induce to definition of ICT focused on their support. Securing of sustainable development of mining industry required definition of ICT, which consider scale and influence of mining production – ICT tools.

## 4      Conclusion

Summarizing, we would like to mark that environmental competency of future mining engineer is a personal formation, which includes the acquired during preparation profession-oriented environmental awareness (cognitive component), adopted ways for securing environmentally safe mining works (praxeological component) in the interest of sustainable development (axiological component) and the qualities of socially responsible ecological behavior (socially-behavioral component) are formed.

Formation of environmental competency is fulfilled during acquirement of the following:

- Socially-personal competences: understanding and perception of ethical norms of behavior in respect to other people and nature (bioethics principals); ecological literacy;
- General scientific competence: deep knowledge in ecology necessary for usage in professional activity;
- Generally professional competence: the ability to use scientific lows and means during evaluation of environmental condition, participate in environmental works, make ecological analysis of events in the field of activity, develop plans on events concerning reduction of manmade load on the environment;
─ Special professional competence: securing of ecologically balanced activity, working knowledge of reasonable and integrated development of geo-recourses potential.

Teaching geoinformation technologies taking into account regional situation, provides personality formation in the natural social and cultural environment. Thus, regional residence differences influence the content of their activities and interests. The education system based on regional characteristics supports interest. That is why teaching should be organized according to actual needs of students. Educational problems should be solved due to initiation and growth of activity, because real environmental situations exist within the students' surrounding.

Thus, solving the problem of ecological competence formation as part of the complex problem in terms of competence approach it requires justifying the choice of using geoinformation technology as a means of ecological competence formation among the future mining engineers.